\documentclass[twocolumn,showpacs,preprintnumbers,amsmath,amssymb]{revtex4}

\usepackage{graphicx}
\usepackage{dcolumn}
\usepackage{bm}

\begin{document}

\preprint{APS-preprint}

\title{Particles held by springs in a linear shear flow exhibit oscillatory motion}

 \author{Lukas Holzer}
 \author{Walter Zimmermann}%
 \email{walter.zimmermann@uni-bayreuth.de}
 \affiliation{%
 Theoretische Physik,
           Universit\"at Bayreuth,
           D-95440 Bayreuth, Germany
 }
 \date{\today}
 \begin{abstract}
The dynamics of small spheres, which are held by linear springs 
in a low Reynolds number shear flow at neighboring locations is investigated. 
The flow elongates the beads and the interplay of the shear 
gradient with the nonlinear behavior of the hydrodynamic interaction among 
the spheres causes in a large range of parameters a bifurcation to a surprising
oscillatory bead motion. 
The parameter ranges, wherein this bifurcation is either super- or subcritical, 
are determined.
 \end{abstract}
\pacs{47.15.G-, 
      47.20.Ky, 
      47.57.J-, 
      47.61.-k 
}
\maketitle
%
%

{\it Introduction.-}
Studies about the motion of bacteria and flagella  in a fluid, 
about the dynamics of blood cells and other small
suspended objects such as polymers in simple flows 
 are currently of central interest 
and count to one of the major issues of microfluidics 
\cite{Brenner:1981,Berg:1993,Purcell:1977.2,Tabeling:2006.1,Chu:99.1,Chu:2005.1,Groisman:2000.1,Berg:2004.1,Groisman:2004.1,Gompper:2005.1,Steinberg:2006.1,Misbah:2006.1,Goldstein:2006.2}. 
According to the short spatial scales involved in these
cases the fluid motion surrounding the particles can be described in the  
small Reynolds numbers limit \cite{Purcell:1977.2,Brenner:1981,Berg:1993,Tabeling:2006.1},
by the linear Stokes equation \cite{Brenner:1981}.

The hydrodynamic interaction ({\bf HI})
between neighboring blood cells, 
swimming bacteria,  different segments of 
a polymer or between neighboring 
polymers is of nonlinear nature even in 
the limit of low Reynolds numbers \cite{Rotne:1969}. Moreover,
this nonlinear behavior may cause dynamical effects, 
such as the periodic motion of small sedimenting spheres  
\cite{Hocking:1964.1,Caflisch:1988.1} 
or the synchronization effects between rotating strings and
between cilia \cite{Stark:2005.1,Vilfan:2006.1}, or it may cause
a hydrodynamic coupling of particles in optical vortices \cite{Grier:2005.1}. The HI
may also amplify thermal fluctuations of polymers \cite{Kienle:2006.1} 
to mention only a few examples. 

Free single polymers show already a complex dynamical behavior 
in shear flow \cite{Chu:99.1,Chu:2005.1}
and the  hydrodynamic interaction between many polymers leads
to the so-called elastic turbulence \cite{Groisman:2000.1,Groisman:2004.1}. 
Furthermore polymers fixed at one end in a plug flow are also 
a major issue \cite{Chu:95.1,Larson:97.1} where
 one finds in this case significant hydrodynamic interaction
effects both for the static as well as dynamic properties of the
tethered polymers \cite{Larson:97.1,Rzehak,Kienle:2006.1}. 
Tethered polymers in shear flow have also been 
studied but so far only a single polymer fixed 
with one end at a wall was considered \cite{Doyle:2000.1,Doyle:2000.2,Gratton:2005.1,Delgado:2006.1}. 
Recently, investigations have been started in order to analyze the 
behavior of several flexible polymers fixed 
with their ends at the top of neighboring
pillars \cite{Bruecker:2005} and exposed
to a linear shear flow.  So the interesting question arises quite 
naturally:  what is the dynamics of
neighboring tethered polymers in shear flow
and which role plays the hydrodynamic interaction?

We mimic a situation of interacting tethered 
polymers by spheres anchored by springs 
and neglect in a first approach
 thermal fluctuations. To the best of our 
knowledge this is the first example where an oscillatory motion of bound 
particles in low Reynolds number 
flow  has its origin in the hydrodynamic interaction.
%
%

%
\begin{figure}[htp]
    \includegraphics[angle=-0,width=.8\columnwidth]{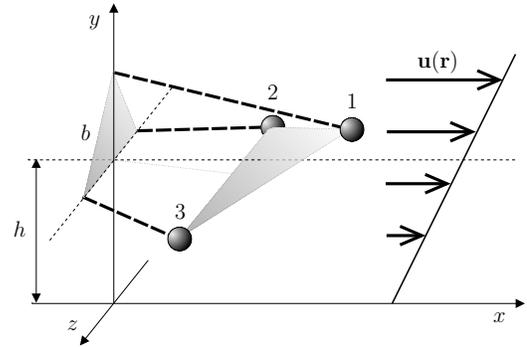}
\vspace{-3mm}
\caption{\label{figconfig}%
The  three particle configuration in shear flow is shown.
The corners of the triangle in the $y-z$-plane mark the minima of 
the harmonic potentials of the beads. 
The flow induced sphere displacements are indicated by 
the dashed lines. 
$h$ measures the shift of the lower side of the triangle from the center of the  shear flow,  ${\bf u}(0)=0$. 
}
\end{figure}
{\it Model.-} Three beads are fixed 
in a shear flow 
by linear springs with a spring constant $k$ as shown
in Fig.~\ref{figconfig}. 
The locations of 
the minima  of the corresponding harmonic potentials, 
 ${\bf R}_i ~(i=1,2,3)$, build, if not stated otherwise,
 an equilateral triangle 
of sidelength $b$ and height $H=b/\sqrt{2}$ with the upper
corner at 
${\bf R}_1=(0,h+H,0)$ and the two lower corners 
at ${\bf R}_{2,3}=(0,h,\pm b/2)$.
The bead-springs are elongated by a linear shear flow 
\begin{equation}
\label{shear}
{\bf u}_0(y)= \left( \dot{\gamma} y,0,0  \right)\,
\end{equation}
and the actual bead positions
${\bf r}_i ~(i=1,2,3)$ are determined by the
equations for the bead velocities
\begin{eqnarray}
\label{e.1}
\dot{{\bf r}}_i &=& {\bf u}_0({\bf r}_i) - \frac{k}{\zeta}\tilde {\bf r}_i + \sum_{j \not =i} \left({\bf u}_{\dot{\gamma}}({\bf r}_{ij}) 
     - \Omega^{RP}({\bf r}_{ij})k\tilde {\bf r}_j\right)
\end{eqnarray}
with $ \tilde {\bf r}_i ={\bf r}_i -{\bf R}_i$. 
The first term describes the linear 
shear flow given by Eq.~(\ref{shear}) and
 the second contribution is the ratio between
the spring force and  
the Stokes friction  $ \zeta=6 \pi \eta a$ that a single fixed particle of effective bead radius $a$ experiences
in a  flow with shear viscosity $\eta$.
Each fixed  particle causes a perturbation of the
shear flow at the location of the other beads and vice versa. This so
called hydrodynamic interaction ({\bf HI}) 
is described for a Stokes flow
by the Rotne Prager tensor \cite{Rotne:1969}
\begin{eqnarray}
\Omega^{RP}({\bf r}) = \frac{1}{8\pi\eta r}
	\left( \left(1 + \frac{2}{3}\frac{a^2}{r^2} \right) \text{I} + \left(1 - 2\frac{a^2}{r^2} \right) \frac{{\bf r}{\bf r}}{r^2}
 	 \right)\,,
\end{eqnarray}
which describes together with the harmonic forces
$k {\tilde {\bf r}}_j$ the fourth contribution in 
Eq.~(\ref{e.1}).  $\text{I}_{ij}=\delta_{ij}$ is 
the unity matrix. The shear flow induces sphere rotations, 
which alter the flow field and therefore its action 
onto other spheres, as described  by the third 
term in Eq.~(\ref{e.1}) \cite{Dhont:1996} 
\begin{eqnarray}
{\bf u}_{\dot{\gamma}}({\bf r}) &=& 
\left( -\frac{5}{2} \left(\frac{a}{r}\right)^3 + \frac{20}{3} \left(\frac{a}{r}\right)^5 \right)
			 \frac{{\bf r}\cdot \text{E} \cdot {\bf r} }{r^2} ~{\bf r} 
   \notag \\
& & \qquad \qquad - \frac{8}{3} \left(\frac{a}{r}\right)^5 \text{E} \cdot {\bf r}\,,
\end{eqnarray}
where
$\text{E}_{ij}=\frac{\dot{\gamma}}{2} (\delta_{ix}\delta_{jy} + \delta_{iy}\delta_{jx})$ 
if the particle can rotate freely
and $\text{E}_{ij}=\dot{\gamma}\delta_{iy}\delta_{jx}$ if an external torque prevents the rotation.
With the relaxation time  $\tau=\zeta/k$ and the effective bead
radius $a$ one may rescale  
time $t \to  \tau t'$, space  ${\bf r} \to a {\bf r}'$ and the 
shear rate $\dot{\gamma}' \to \tau \dot {\gamma}$
and the results in this work are most conveniently presented in terms of these
dimensionless units, as for instance 
the sphere displacement 
$
{\bf r}_{d;i}= {\bf r}'_i-{\bf R}'_i =\left(x_{d;i},y_{d;i},z_{d;i}\right)\,.
$
%
%
\begin{figure}
    \includegraphics[angle=-0,width=.9\columnwidth]{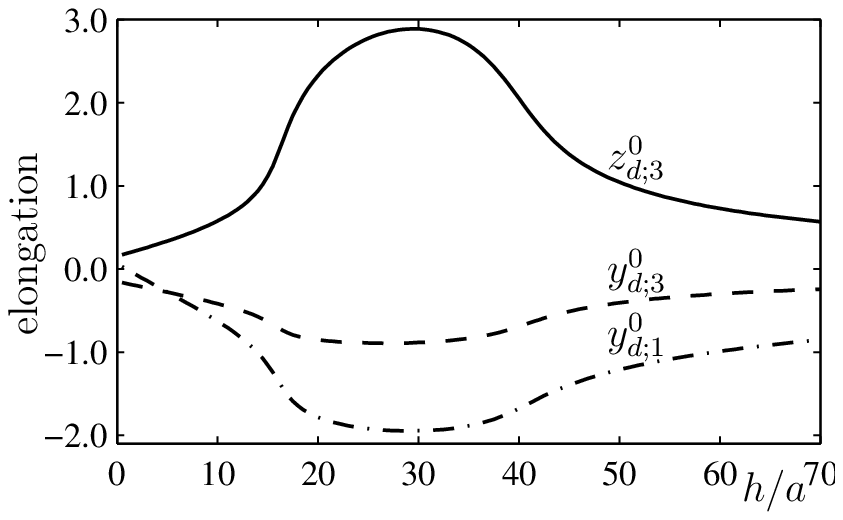}
\vspace{-3mm}
\caption{\label{figdisp}%
The figure shows the stationary vertical 
displacements   $y^0_{3}=y^0_{2}$ of the lower beads
 (dashed line) and $y^0_1$ of the upper bead (dash-dotted line) 
as well as the shift  $z^0_{3}=-z^0_{2} $ in $z$-direction 
(solid line) as a function of the vertical shift $h/a$
and for the dimensionless shear rate $\tau \dot{ \gamma } =2.6$.
}
\end{figure}
%
%
%

{\it Stationary displacement of the spheres.}-
The stationary solutions of the nonlinear 
equations (\ref{e.1}),  
i.e. $\dot {\bf r}_i=0$, with the displacements
${\bf r}^0_{d;i}:={\bf r}'^0_{i}- {\bf R}'_{i}
=\big(x^0_{d;i},y^0_{d;i},z^0_{d;i}\big)$
are determined numerically by a
Newton algorithm. 
 ${\bf r}^0_{d;i}$ of a single bead
increases according to the Stokes drag 
force ${\bf F}=6 \pi \eta a {\bf u}$
and the linear spring force
linearly with the flow velocity  ${\bf u}_0$. 
 However, by virtue of the nonlinear nature of
the HI between the beads 
the  elongation of the linear springs 
change nonlinearly as a function of the flow velocity, which itself varies
linearly with the height $h/a$, as depicted 
in Fig.~\ref{figdisp} for $\tau\dot{ \gamma } =2.6$.

For $h=0$ the flow velocity vanishes at the positions of
bead  $2$ and $3$ and with a finite shear gradient
$\dot \gamma$ only the upper bead
$1$ is displaced. The flow perturbation caused by bead $1$  
shifts beads $2$ and $3$ slightly downward and pushes 
both away in $z-$direction with $z_{d;3}^0=-z_{d;2}^0$.
 For finite  $h$ bead $2$ and $3$ are exposed 
to a finite velocity  ${\bf u}_0({\bf r}_{2,3})$
and excite  also  flow perturbations pointing 
both downward at bead $1$. 
This involves  $y^0_{d;1}$ to become negative as well
and since both perturbations act 
downwards one has  $y^0_{d;1}< y^0_{d;2}$
at intermediate
values of $h$, cf. Fig.~\ref{figdisp}.
In  $z$-direction both disturbances 
compensate each other, so that $z_{d;1}^0=0$
is left unchanged. 
According to this stronger displacement  $y^0_{d;1}$ 
at intermediate values of $h$ the relative distance
between the upper and the two lower beads is reduced and therefore
the flow perturbations caused by bead $1$ are enhanced, and so are
the values of $|y_{d;3}^0|$ and $z_{d;3}^0=-z_{d;2}^0$ as a function 
of $h$.

It is  very surprising that all these displacements 
reach extrema, as shown in Fig.~\ref{figdisp} and become smaller 
again for large values of $h$.  An explanation of this
behavior may be offered by inspecting the bead positions
at large values of $h/a$. In this case the triangle 
built by the bead positions is again nearly parallel
to the $y-z$ plane and accordingly the flow perturbations and
the related forces 
 caused at the neighboring beads are 
nearly vanishing compared to their external force. 
In this limit, however, the height $H$ of the 
triangle is smaller and the distance between the beads $2$
and $3$ is larger than for a vanishing fluid velocity.
The latter behavior is a consequence of a complex balance 
between the  spring forces and the nonlinear forces due to the flow 
disturbances. Correspondingly there is hitherto no simple 
qualitative picture for both, the deformed triangle built by the
beads and the decreasing behavior of the elongation beyond
their extrema.
 \begin{figure}[htb]
    \includegraphics[angle=-0,width=.85\columnwidth]{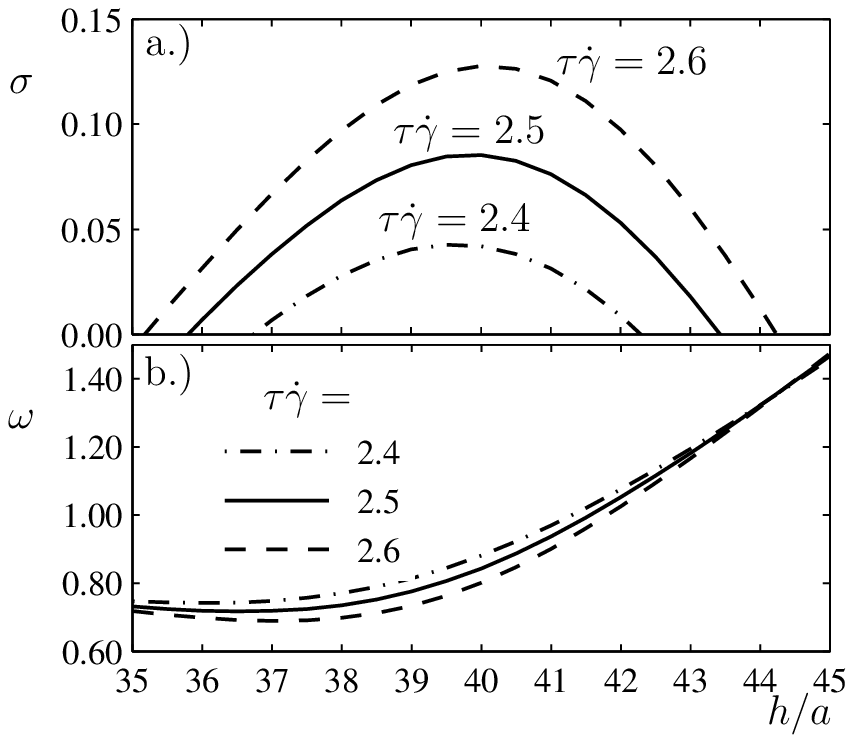}
\vspace{-3mm}
\caption{\label{figomega}%
The largest real part $\sigma(h)$ is given in a)  as a function
of the vertical shift $h/a$ 
 for different shear rates $\dot{\gamma}$
and part b) shows the corresponding 
imaginary parts $\omega$. 
 }
\end{figure}
%
%
%

\begin{figure}[htb] 
\centering
    \includegraphics[angle=-0,width=.9\columnwidth]{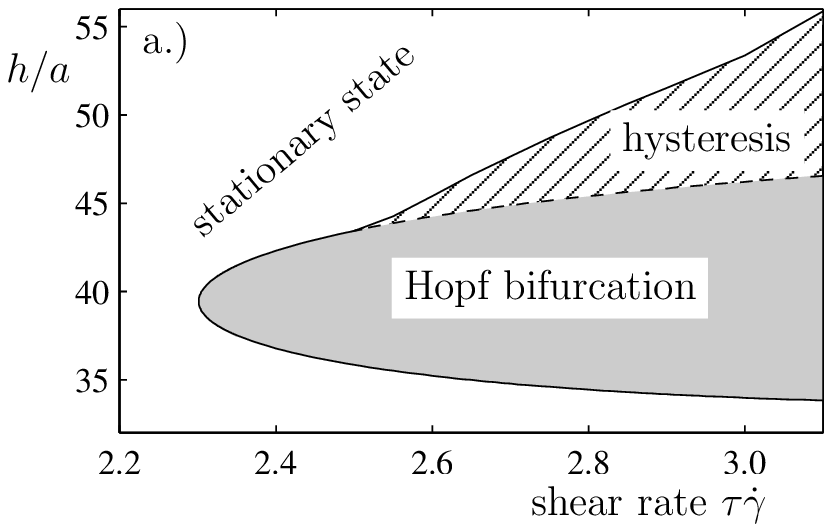}
    \includegraphics[angle=-0,width=.9\columnwidth]{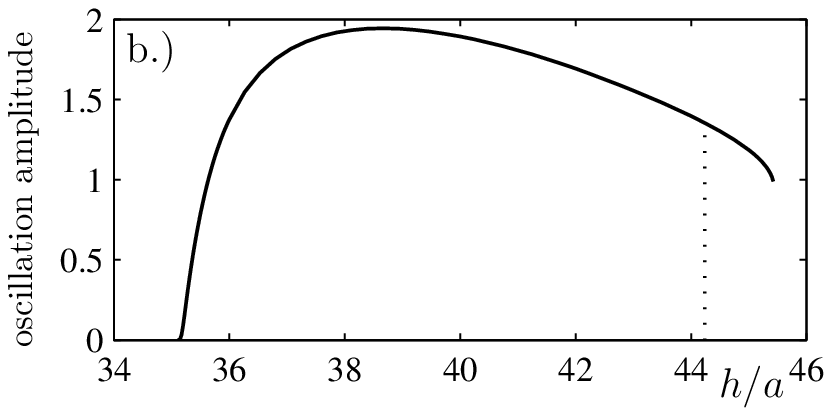}
\vspace{-3mm}
\caption{\label{figlinstab}%
a.) In the grey  range the
stationary bead elongation are  unstable with respect 
to a supercritical Hopf bifurcation along the solid border line and to a subcritical one along the dashed line. 
Within the striped region the Hopf bifurcation is hysteretic. 
b.) The oscillation amplitude of bead $1$ in $x$-direction is given
as a function of $h$ for $\tau \dot{\gamma} =2.6$.
The dotted line marks the upper threshold of the Hopf bifurcation.
}
\end{figure}

{\it Threshold of the Hopf bifurcation.-}
Slightly beyond the extrema
 in  Fig.~\ref{figdisp} the stationary 
bead displacements become unstable and 
one finds by numerically integrating 
Eqs.~(\ref{e.1}) using a standard method a bifurcation to
oscillatory bead motions. The threshold of 
this bifurcation may be determined by a linear 
stability analysis of  the stationary elongation
${\bf r}^0_{d;i}$ with respect to small perturbations
  $\delta{\bf r}_{i}(t)$.
Using the ansatz 
  $ {\bf r}_{i}={\bf r}^0_{i} +\delta{\bf r}_{i}(t)$,
a linearization of Eqs.~(\ref{e.1}) leads to a set of 
$9$ linear differential equations with constant coefficients 
\begin{eqnarray}
\label{linstab}
\dot {\bf Y} = {\cal L}({\bf r}_i^0) {\bf Y} \qquad {\rm with}\quad
{\bf Y}(t) = \left( \delta{\bf r}_{1},\delta{\bf r}_{2},\delta{\bf r}_{3}\right)\,,
\end{eqnarray}
governing the linear dynamics 
of the  perturbations  $\delta{\bf r}_{i}(t)$. 
Eq.~(\ref{linstab}) is solved by  
${\bf Y} =\exp(\sigma t' \pm i \omega t') {\bf Y}_0$, which transforms 
Eq.~(\ref{linstab}) into
an eigenvalue problem.
The eigenvalue  with the largest real part $\sigma(h)$ has  also a finite 
imaginary part $\omega$ and is positive within a
finite range of  $h/a$ as shown in 
in Fig.~\ref{figomega}
for three different values 
of the dimensionless shear rate $\tau\dot{\gamma} $.  
The whole range of a positive $\sigma(h)$ 
in the $\tau \dot{ \gamma } - h/a$ plane is given by
the shaded range in Fig.~\ref{figlinstab}. 
The occurrence of oscillatory bead motion is rather robust with
respect to changes of the anchor points of the linear
springs which adhere the beads. We have tested this by changing
the anchor point of bead $1$ and $2$ in all three
spatial directions.  With such modifications
the three anchor points build either no equilateral 
triangle or the equilateral triangle is inclined and not
anymore perpendicular to the flow direction. The major
trends include the following ones. 
Bringing anchor points closer together enhances the hydrodynamic
interaction which favors the Hopf bifurcation in a larger
parameter range and it then 
takes also place at smaller shear rates and $h$.
\begin{figure}[ht]
        \includegraphics[angle=-0,width=.9\columnwidth]{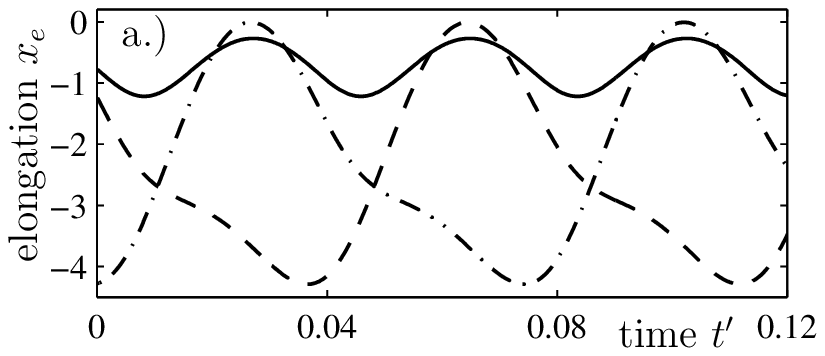}
    \includegraphics[angle=-0,width=.9\columnwidth]{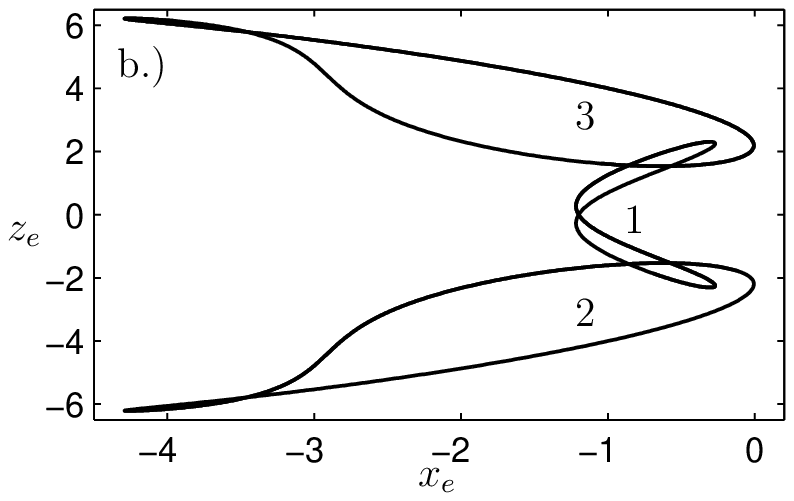}
\vspace{-3mm}
\caption{\label{beadosz}%
a.) The time dependence of the
deviation ${\bf r}_{e;i}$  from the center of mass for
 bead $1$ (solid line), $2$ (dashed line)
and $3$ (dash-dotted line) for 
$\tau\dot{\gamma}=2.6$ and $h/a=35.5$.
b.) The corresponding bead oscillations 
in the $x-z$ plane.
}
\end{figure}
%
%
%

{\it Nonlinear behavior of the bead oscillations.-}
A typical example for the three dimensional oscillatory motion of 
the beads is given by a projection
onto the $x$-axis in Fig.~\ref{beadosz}a). 
Here the deviations ${\bf r}_{e,i}=(x_{e;i},y_{e;i},z_{e;i})$   
from the center of mass of the stationary solutions 
${\bf r}_{cm}=
\sum_{i=1,2,3} {\bf r}^0_i/3$ are displayed. 
Two characteristic features can be recognized. 
Bead $2$ and $3$ oscillate with a phase shift of $\pi$ and bead $1$ 
oscillates along the $x$-direction with twice the frequency 
of the  other two  beads. The double frequency of bead $1$ 
 is an effect  of the projection onto the $x$-axis, as can be
seen from the phase portrait in  Fig.~\ref{beadosz}b).
Similar phase portraits can be  obtained in the $x-y$ and the 
$y-z$ plane as well. Bead $1$  performs a three dimensional
motion and accordingly beads $2$ and $3$ are
pushed away by a phase shift of $\pi$ as indicated in Fig.~\ref{beadosz}a).

The Hopf bifurcation is supercritical along 
the solid line bounding the grey 
range in Fig.~\ref{figlinstab}a). It is subcritical 
along the dashed one and the range
of hysteresis is indicated by the striped range.
The  oscillation amplitude $\delta {\bf r}_i^2$ 
of bead $1$ is
shown in  Fig.~\ref{figlinstab}b) as a function of  $h/a$ 
at the shear rate $\tau\dot{\gamma}=2.6$. It indicates 
the supercritical behavior at the lower threshold and
the hysteresis at the upper one.
Close to the supercritical  Hopf bifurcation
the oscillations are  harmonic. Further
away from this threshold and in the parameter
range with hysteresis in Fig.~\ref{figlinstab}a)
the periodic motion becomes rather anharmonic.
%
%

{\it Conclusions.-} We found in this work a
Hopf bifurcation of three bounded spheres  in a 
low Reynolds number linear
shear flow, which is induced by the interplay of
the nonlinear behavior of  hydrodynamic interaction 
between the spheres and  the shear gradient.
To the best of our knowledge it is the first description
of oscillations of bounded and hydrodynamically interacting 
particles in a Stokes flow.  Most of the results are obtained
for three beads anchored by linear springs
at the corner of an equilateral 
triangle which is perpendicularly oriented with respect
to the flow direction. The phenomenon is very 
robust against various variations of the anchor points.
For two beads we did not find oscillations.

Our results may also  guide 
investigations on hydrodynamically interacting 
polymers fixed at small spheres and held by laser tweezers or
anchored at  boundaries in shear flow as well as for polymers 
that are fixed in shear flow  close to boundaries
at the top of pillars \cite{Bruecker:2005}. It is also 
an interesting question to be addressed, whether a recently
discussed cyclic motion for grafted polymers 
\cite{Doyle:2000.1,Gratton:2005.1,Delgado:2006.1}
is related to the Hopf bifurcation 
discussed here.

We expect  that
several modifications of our model favor oscillatory motion too.
For instance 
nonlinear spring constants (which may mimic tethered polymers), 
different spring constants
in different directions  or an exposition of the three beads 
to a Poiseuille flow with its spatially dependent shear rate. 
The effects of these and other extensions are the 
subject of forthcoming work.

%
%
\acknowledgements 
We are grateful to J.~Bammert, R.~Peter and F.~Ziebert for useful discussions.
We  acknowledge financial support from the German science foundation (DFG)
via the priority program SPP 1164.

\end{document}